\documentclass[useAMS,usenatbib]{mn2e}
\usepackage{graphicx}
\usepackage{color}
\usepackage{ulem}

\usepackage{epsf}
\def\msun{{\rm M_{\odot}}}

\def\me{{\dot M_{\rm Edd}}}
\def\le{{L_{\rm Edd}}}
\def\mtr{{\dot M_{\rm tr}}}

\title[ULXs: Neutron Stars vs Black Holes] {ULXs: Neutron Stars vs Black Holes}

\author[Andrew King and Jean--Pierre Lasota]
{Andrew King$^{1, 2}$ and Jean--Pierre Lasota$^{3,4}$\\
$^{1}$ Theoretical Astrophysics Group, Department of Physics \& Astronomy, University of Leicester, Leicester LE1 7RH, UK\\
$^{2}$ Astronomical Institute Anton Pannekoek, University of Amsterdam, Science Park 904, 1098 XH Amsterdam, Netherlands\\
$^{3}$ Institut d'Astrophysique de Paris, CNRS et Sorbonne Universit\'es, UPMC Paris~06, UMR 7095, 98bis Bd Arago, 75014 Paris, France\\
$^{4}$ Nicolaus Copernicus Astronomical Center, Bartycka 18, 00-716 Warsaw, Poland     
}

\date{\today}

\volume{000}

\pagerange{\pageref{firstpage}--\pageref{lastpage}} \pubyear{2014}

\begin{document}

\label{firstpage}

\maketitle

\begin{abstract}
We consider ultraluminous X--ray sources (ULXs) where the accretor is a neutron star rather than a black hole. We show that the recently--discovered example (M82 X--2) fits naturally into the simple picture of ULXs as beamed X--ray sources fed at super--Eddington rates, provided that its magnetic field is weaker ($\simeq 10^{11}{\rm G}$) than a new--born X--ray pulsar, as expected if there has been mass gain. Continuing accretion is likely to weaken the field to the point that pulsing stops, and make the system indistinguishable from a ULX containing a black hole. Accordingly we suggest that a significant fraction of all ULXs may actually contain neutron star accretors rather than black holes, reflecting the neutron--star fraction among their X--ray binary progenitors. We emphasize that neutron--star ULXs are likely to have {\it higher} apparent luminosities than black hole ULXs for a given mass transfer rate, as their tighter beaming outweighs their lower Eddington luminosities. This further increases the likely proportion of neutron--star accretors among all ULXs. Cygnus X--2 is probably a typical descendant of neutron--star ULXs, which may therefore ultimately end as millisecond pulsar binaries with massive white dwarf companions.
\end{abstract}

\begin{keywords}
  accretion, accretion discs -- binaries: close -- X-rays: binaries --
  black hole physics -- neutron stars -- pulsars: general  
 \end{keywords}

\section{Introduction}

Ultraluminous X--ray sources (ULXs) have apparent luminosities $L$ above the Eddington limit for a standard stellar--mass accretor (often taken as $L \ga 10^{39}\,{\rm erg\, s^{-1}}$), but are not in the nuclei of their host galaxies. These defining properties initially led to suggestions that their accretors were black holes with masses higher than stellar, but below supermassive -- called intermediate--mass black holes \citep[IMBH:][]{CM99}. It is now widely accepted instead that most -- or perhaps all -- ULXs are standard X--ray binaries in unusual and shortlived phases  \citep{Kingetal01}. 

Perhaps because the required breach of the Eddington limit is then minimised, many authors have assumed that the accretor in ULXs is always a black hole. But in a striking recent paper \citet{Bachettietal14} find a ULX  (M82 X--2) with $L \simeq 1.8\times10^{40}\,{\rm erg\,s^{-1}}$ (about 100 times Eddington) and a coherent periodicity $P = 1.37$~s, which they interpret as the spin period of an accreting magnetic neutron star (X--ray pulsar). One line of explanation of this result \citep[cf][]{Tong15,Eksi15,Dosso15} uses the fact that the electron--scattering cross--section is significantly reduced (by a factor  ($\sim (\nu/\nu_B)^2$) for frequencies $\nu$ below cyclotron ($\nu_B$)) for certain polarizations and field directions, allowing luminosities which exceed the usual Eddington limit by factors $\sim (\nu_B/\nu)^2$. The effect is greatest for the strongest fields, as found in magnetars. In this picture M82 X--2 would evidently be completely different from all other ULXs found so far. 

In this paper we adopt a very different viewpoint, and try to see if M82 X--2 instead fits naturally into a unified picture applying to all ULXs. This is promising, since there is a lot of evidence that the basic cause of the unusual behaviour of ULXs is a super--Eddington mass supply, and this is evidently perfectly possible for a neutron star accretor (see below).
So we assume that for reasons discussed below, the binary supplies mass to the vicinity of the accretor in ULXs (including M82 X--2) at a rate significantly above the value $\me$ that would produce the Eddington luminosity $\le$ if all of it could be accreted. Much of this mass is instead driven off by radiation pressure,  producing geometrical collimation or beaming of most of the emission (see Section 2 below). As a result we generally detect only ULXs beamed towards us. Multiplying their fluxes by $4\pi D^2$, where $D$ is the source distance, then overestimates the intrinsic source luminosity by a factor $1/b \gg 1$, where $b = \Omega/4\pi \ll 1$ is the beaming factor ($\Omega$ is the true solid angle of emission). 

In this picture ULX behaviour requires only that the mass supply rate is $\gg \me$, whatever the nature of the accretor defining $\me$. It follows \citep{Kingetal01} that {\it ULX accretors can be neutron stars or white dwarfs} provided only that the supply rate exceeds
$10^{-8}\msun\,{\rm yr^{-1}}$ or $ 10^{-5}\msun\,{\rm yr^{-1}}$ respectively. In line with this prediction \citet{Fabbianoetal03} suggested that one of the ULXs in the Antennae probably has a white dwarf accretor. Similarly, the magnetic neutron star in M82 X--2 can appear as a ULX provided that mass is supplied at a rate $\gg 10^{-8}\msun\,{\rm yr^{-1}}$.

Our aim here then is to see how pulsed systems like M82 X--2 fit into a general picture of ULXs as binaries with super--Eddington mass supply rates and beamed emission. The interesting aspect is that the high observed spinup rate of M82 X--2 gives information about the magnetic accretion process. \citet[][hereafter KL]{KL} make this point: they
investigated M82 X--2 as a system containing a weak--field pulsar, but their model does not explain the observed super--Eddington apparent luminosity. We will find that to make the observed spinup rate consistent with the beaming needed to produce the apparent luminosity determines the magnetic moment of the neutron star. The result implies that the beaming process works at accretion disc radial scales which are larger than those where magnetic effects become important.
M82 X--2 is then effectively a fairly normal accreting magnetic neutron star inside a collimating disc structure arising from the super--Eddington mass supply. 

\section{ULX Accretion}

There are (at least) two classes of ULXs, corresponding to super--Eddington mass supply in two distinct situations (King, 2002): 

(a) thermal--timescale mass transfer in high--mass X--ray binaries (HMXBs), which is the natural sequel to the usual HMXB wind--capture phase once the companion fills its Roche lobe, and 

(b) long-lasting transient accretion disc outbursts in low--mass X-ray binaries. 

M82 X--2 is clearly in the first group (cf KL) as it is known (Bachetti et al. 2014) to have a stellar companion of mass $M_2\ga 5.2\msun$ and radius $R_2\ga 7\msun$. This star must be filling its Roche lobe, given the binary period of 2.5\,d. Since $M_2$ is significantly larger than the likely neutron star mass, the binary and the companion's Roche lobe must be shrinking because mass is transferred to the neutron star orbit, further from the binary's centre of mass. Mass transfer on the companion's thermal timescale results \citep[cf][]{KR99,KB99, Kingetal00,PR00}, and can give rates as high as $\mtr \sim 10^{-5}\msun\,{\rm yr^{-1}}$ depending on the companion mass and its degree of lobe--filling. We will find that M82 X--2 probably has a more modest (but still strongly super--Eddington) transfer rate $\mtr \sim 1.2\times 10^{-6}\msun\,{\rm yr^{-1}}$, suggesting that it is fairly near the beginning of the thermal--timescale phase \citep[cf][]{KR99}
characterizing its ULX stage. We will see that this is probably the reason that this system pulses at all.

Given a super--Eddington mass supply, the resulting accretion disc is stable at large disc radii, down to the `spherization radius'
\begin{equation}
R_{\rm sph} \simeq \frac{27}{4}\dot m_0 R_g
\label{rsph}
\end{equation}
\citep{SS73}, where the local energy release is close to the local Eddington value and there must be significant outflow (here $R_g = GM/c^2 =2.1\,{\rm km}$ is the gravitational radius of the neutron star, of mass $M \sim 1.4\msun$, and $\dot m_0$ is the Eddington factor far from the accretor, so for thermal--timescale systems $\dot m_0 = \mtr/\me$). We will find that for M82 X--2, $R_{\rm sph}$ is  larger than the 
magnetospheric radius $R_M$ (see eqn \ref{rmf} below) where magnetic effects become important. 

We expect outflow from disc radii inside $R_{\rm sph}$ also. To keep each
disc radius close to its local Eddington limit, the outflow must ensure that the accretion rate at disc radius R decreases as
\begin{equation}
\dot M(R) \simeq \frac{R}{R_{\rm in}}\me
\label{mr}
\end{equation}
where $R_{\rm in} \sim R_M$ is the innermost disc radius.
Integrating the local disc emission shows \citep[cf][]{SS73} that the total accretion luminosity is
\begin{equation}
L \simeq \le\left[1 + \ln\dot m_0\right].
\label{ldisc}
\end{equation}

The main constraint on $\dot m_0$ for M82 X--2 comes from the beaming factor $b$, which has to account for most of the difference between its apparent luminosity $\simeq 10^{40}\, {\rm erg\, s^{-1}}$  and the Eddington luminosity $\le \simeq 2\times 10^{38}\, {\rm erg\, s^{-1}}$ of a $1.4\msun$ neutron star. \citet{King09} gives an approximate formula
\begin{equation}
b \simeq \frac{73}{{\dot m_0}^2}
\label{beam}
\end{equation}
valid for $\dot m_0 \ga \sqrt{73} \simeq 8.5$.
This form ensures that soft X--ray components in ULX spectra \citep[cf][]{KajavaP09} obey an inverse luminosity--temperature correlation $L_{\rm soft}\sim T^{-4}$, as observed. The $b \propto {\dot m_0}^{-2}$ scaling (but not its normalization) also follows quite independently by noting (cf the discussion before eqn (\ref{ldisc}) above) that the vertical
size of the disc structure near $R_{\rm sph}$ must scale with $\dot m_0$\footnote{In black--hole accretion the innermost part of a super--Eddington flow is advection dominated and  the height of the accretion flow is independent of the accretion rate \citep{Oleketal15, meOlek15}; it is not clear what happens in accretion on to a neutron star.}, while all the central accretors essentially gain mass at their Eddington rate ($\dot m = 1$) and so are self--similar. This form of beaming is found to reproduce the local luminosity function of ULXs very well \citep{Mainierietal10}. 

Combining (\ref{ldisc}, \ref{beam}) to give the apparent (spherical) luminosity $L_{\rm sph} = L/b$ we find 
\begin{equation}
\frac{m_1}{L_{40}} \simeq \frac{4500}{{\dot m_0}^2(1 + \ln\dot m_0)}
\label{mL}
\end{equation}
\citep{King09} where $m_1$ is the accretor mass in $\msun$ and $L_{40}$ is the apparent luminosity in units of $10^{40}\, {\rm erg\, s^{-1}}$. For M82 X--2 we take
$m_1 = 1.4, L_{40} = 1.8 $, which gives 
\begin{equation}
\dot m_0\simeq 36
\label{M0}
\end{equation}
fixing the mass transfer rate in the binary as 
\begin{equation}
\dot M_0 \simeq 1.2 \times 10^{-6}\msun\,{\rm yr}^{-1}.
\label{dotM0}
\end{equation}
We will see in the next Section that mass accretes near the magnetosphere at a rate significantly smaller than this, self--consistently implying from (\ref{mr}) that the magnetospheric radius $R_M$ (eqn \ref{rm}) is smaller than the spherization radius 
\begin{equation}
R_{\rm sph} \simeq 5 \times 10^7\,{\rm cm},
\label{rsphnum}
\end{equation}
defining the ULX beaming.

\section{Spin and Spinup of M82 X--2}

Magnetic accretors are characterized observationally by their spin period $P$ and spinup rate
$\dot \nu = {\rm d}(2\pi/P)/{\rm d}t$. For M82 X--2 we have $P = 1.37$~s, and an unusually high value $\dot\nu = 2\times 10^{-10}\, {\rm s^{-2}}$. This gives a very short spinup (period halving) timescale 
\begin{equation}
t_{\rm spin} \simeq \frac{2\pi}{P\dot\nu} \simeq 360\,{\rm yr}.
\label{spinup}
\end{equation}
Physically, the magnetic accretion process is characterized by the magnetospheric (`Alfv\'en') radius $R_M$ where the matter stresses in the accretion disc are comparable with those of the magnetic field of dipole moment $\mu = 10^{30}\mu_{30}\,{\rm G\,cm^{-3}}$, i.e.
\begin{equation}
R_M = 2.9\times 10^8 \dot M_{17}^{-2/7}m_1^{1/7}\mu_{30}^{4/7}\,{\rm cm},
\label{rm}
\end{equation}
where $\dot M_{17}$ is the accretion rate at the magnetosphere in units of $10^{17}\,{\rm g\,s^{-1}}$ \citep[cf e.g.][]{FKR02}. Disc accretion is assumed to give way to flow along fieldlines within $R_M$, so the disc angular momentum arriving at $R_M$ predicts a theoretical spinup rate 
\begin{equation}
\dot\nu = 2.7\times 10^{-12}\dot M_{17}^{6/7}m_1^{-3/7}R_6^{6/7}\mu_{30}^{2/7}I_{45}^{-1}{\rm Hz\,s^{-1}}
\label{nudot}
\end{equation}
where $R_6$ is the neutron star radius in units of $10^6$\,cm,  and $I_{45}$ its moment of inertia in units of $10^{45}{\rm \,g\,cm^2}$ \citep[see e.g.][]{FKR02}.

Using the observed value of $\dot\nu$ in this equation and taking $m_1=1.4$, $R_6=I_{45}=1$ gives the current value of the local accretion rate at $R = R_M$ as 
\begin{equation}
\dot M(R_M) \simeq 2.8\times 10^{-7}\mu_{30}^{-1/3}\,{\msun\,{\rm yr}^{-1}}.
\label{mdotrm}
\end{equation}
For self--consistency we must now impose the scaling (\ref{mr}), in the form
\begin{equation}
\frac{R_{\rm sph}}{R_M} = \frac{\mtr}{\dot M(R_M)},
\label{consist}
\end{equation}
which gives
\begin{equation}
\mu_{30} \simeq 0.1.
\label{mu}
\end{equation}
The magnetospheric radius is then
\begin{equation}
R_M \simeq 2.0 \times 10^7\,{\rm cm},
\label{rmf}
\end{equation}
and the accretion rate at the magnetospheric radius $\dot M(R_M) \simeq 2\times 10^{-7}\,{\msun\,{\rm yr}^{-1}}$.

\section{Evolution}

Since $\mu = BR_*^3$ we see from (\ref{mu}) that the magnetic field of the neutron star in M82 X--2 must be significantly lower than is usual for a new--born neutron star, i.e. $B \simeq 10^{11}\, {\rm G}$, rather than $\simeq 10^{12}\,{\rm G}$. This is reasonable, since it has long been suspected that accretion of even a relatively small mass severely reduces the surface fields of neutron stars -- this is central to the concept of pulsar recycling, which is implicated in the production of millisecond pulsars \citep{RadSri82,Alparetal82,Taamvdh86,BhavdH91}. It is unclear whether M82 X--2 or systems like it will end by producing millisecond pulsars or not when accretion stops: given its spin period and deduced magnetic field, the neutron star in M82 X--2 is currently below the pulsar `death line'
\begin{equation}
BP^{-2} \simeq 1.7\times 10^{11}\,{\rm G\, s^{-2}}
\label{death}
\end{equation}
\citep{RS75}, so resurrection as a pulsar will require spinup to beat field decay. We can expect M82 X--2 to transfer several $\msun$ towards its neutron--star companion. But since the surface accretion rate is limited to $\me$ or even less (perhaps $\me$ multiplied by the fractional surface area of the accreting polecaps) the neutron--star mass grows much less. A clear example of this is Cyg X--2, where a prolonged super--Eddington phase has nonetheless left behind a fairly normal (but not strongly magnetic) neutron star, which may yet live again as a radio pulsar once accretion stops \citep{KR99}. 

\section{Conclusion}

We have seen that M82 X--2 fits into the simple picture of ULXs as beamed X--ray sources fed at super--Eddington rates. Its magnetic field has apparently been weakened by accretion, and we can expect that its field will shortly be unable to channel the flow (formally $R_M < R_*$) so that pulsing will stop. The system will then be indistinguishable from a ULX containing a black hole. The mass transfer rate from the companion will increase substantially \citep[see][]{KR99} so the current ULX might become a HLX (hyperluminous X-ray source).

On this basis we suggest that a significant fraction of {\it all} ULXs might actually contain neutron star accretors rather than black holes. This is already plausible because a large fraction of the high--mass X--ray binary progenitors of ULXs have such accretors, and 
in principle dynamical mass measurements might offer a way of checking this idea.
Perhaps surprisingly, neutron--star ULXs actually have {\it higher} apparent luminosities than black--hole ULXs for a given mass supply rate $\dot M_{\rm supp}$ (cf King, 2006). Using $\dot m_0 = \dot M_{\rm supp}/\me \propto 1/m_1$ we can rewrite (\ref{mL}) as 
\begin{equation}
L \propto \frac{C - \ln m_1}{m_1},
\label{nsbh}
\end{equation}
with $C$ a constant. In other words for a lower accretor mass, the tighter beaming resulting from a higher Eddington accretion ratio outweighs the decrease in the Eddington luminosity. In most cases -- certainly the thermal--timescale mass transfer discussed here -- $\dot M_{\rm supp}$ is independent of $m_1$. Then we expect neutron--star ULXs to have higher apparent luminosities than black--hole ones and so to be relatively easier to find, further increasing the number of neutron--star accretors among detected ULXs. A constraint on the relative number of neutron--star versus black--hole ULXs produced by this `early massive Case B evolution' comes from the
delayed dynamical instability \citep{W77,Hjel89}, which probably requires initial mass ratios $M_{2i}/M_{1i} \la 4$ to avoid the binary merging in a common--envelope system.
Evidently a population synthesis calculation starting from the observed ratio of neutron--star to black--hole systems is needed to give a good idea of this ratio for their ULX descendants \citep[see also][]{Fragos15,Wiktor15,YL15}.

The ULX phase (initially pulsed, but later unpulsed) ends once the mass transfer rate drops below Eddington, and for neutron--star systems like M82 X--2 will probably give a system resembling Cyg X--2. Here an apparently nonmagnetic neutron star accretes at modest rates from an extended but low--mass companion star, which is significantly hotter than expected for a low--mass giant. It is worth noting that the neutron star has evidently gained very little of the mass 
lost by the initially massive companion: it accretes only at its own Eddington limit rate ($\dot M_1 \sim 10^{-8}\msun\,{\rm yr}^{-1}$),  and mass
transfer lasts for of order the initial thermal timescale ($\sim 10^5 - 10^6\,{\rm yr}$) of the companion.  As discussed by \citet{KR99}, the end products of this evolution include binary millisecond pulsars in wide binaries with relatively massive white dwarf companions, and systems with much shorter periods ($\la 1\,{\rm day}$).

 \section*{Acknowledgments}

JPL acknowledges support from the French Space Agency CNES and Polish NCN grants UMO-2013/08/A/ST9/00795 and DEC-2012/04/A/ST9/00083. ARK thanks the Institut d'Astrophysique, Paris, for hospitality during a visit where this work was performed. Theoretical astrophysics research at the University of Leicester is supported by an STFC Consolidated Grant.

\bsp

\label{lastpage}


\begin{thebibliography}{99}

\bibitem[Alpar et al.(1982)]{Alparetal82} Alpar, M.~A., Cheng, 
A.~F., Ruderman, M.~A., \& Shaham, ~J.\ 1982, Nature, 300, 728

\bibitem[\protect\citeauthoryear{Bachetti et 
al.}{2014}]{Bachettietal14} Bachetti M., et al., 2014, Nature, 514, 202 

\bibitem[\protect\citeauthoryear{Begelman, King, 
\& Pringle}{2006}]{Begelman06} Begelman M.~C., King A.~R., Pringle J.~E., 2006, MNRAS, 370, 399 

\bibitem[Bhattacharya 
\& van den Heuvel(1991)]{BhavdH91} Bhattacharya, D., \& van den Heuvel, E.~P.~J.\ 1991, Phys. Rep., 203, 1 

\bibitem[\protect\citeauthoryear{Colbert 
\& Mushotzky}{1999}]{CM99} Colbert E.~J.~M., Mushotzky R.~F., 1999, ApJ, 519, 89 

\bibitem[Dall'Osso et al.(2015)]{Dosso15} Dall'Osso, S., Perna, 
R., Papitto, A., Bozzo, E., \& Stella, L.\ 2015, arXiv:1512.01532 

\bibitem[Ek{\c s}i et al.(2015)]{Eksi15} Ek{\c s}i, K.~Y., 
Anda{\c c}, {\.I}.~C., {\c C}{\i}k{\i}nto{\u g}lu, S., et al.\ 2015, 
MNRAS, 448, L40 

\bibitem[\protect\citeauthoryear{Fabbiano et 
al.}{2003}]{Fabbianoetal03} Fabbiano G., King A.~R., Zezas A., Ponman 
T.~J., Rots A., Schweizer F., 2003, ApJ, 591, 843 

\bibitem[Frank et al.(2002)]{FKR02} Frank, J., King, A., 
\& Raine, D.~J.\ 2002, Accretion Power in Astrophysics, by Juhan Frank and Andrew King and Derek Raine, ~Cambridge, UK: Cambridge University Press

\bibitem[Fragos et al.(2015)]{Fragos15} Fragos, T., Linden, T., 
Kalogera, V., \& Sklias, P.\ 2015, ApJL, 802, L5 

\bibitem[\protect\citeauthoryear{Hjellming}{1989}]{Hjel89}Hjellming, M.S., 1989, PhD thesis, University of Illinois

\bibitem[\protect\citeauthoryear{Kajava 
\& Poutanen}{2009}]{KajavaP09} Kajava J.~J.~E., Poutanen J., 2009, MNRAS, 398, 1450 

\bibitem[\protect\citeauthoryear{King}{2002}]{King02} King 
A.~R., 2002, MNRAS, 335, L13 

\bibitem[\protect\citeauthoryear{King}{2009}]{King09} King 
A.~R., 2009, MNRAS, 393, L41 

\bibitem[\protect\citeauthoryear{King}{2014}]{King14} King A., 
2014, Sci, 343, 1318 

\bibitem[\protect\citeauthoryear{King 
\& Begelman}{1999}]{KB99} King A.~R., Begelman M.~C., 1999, ApJ, 519, L169 

\bibitem[\protect\citeauthoryear{King et al.}{2001}]{Kingetal01} 
King A.~R., Davies M.~B., Ward M.~J., Fabbiano G., Elvis M., 2001, ApJ, 
552, L109 

\bibitem[\protect\citeauthoryear{King 
\& Ritter}{1999}]{KR99} King A.~R., Ritter H., 1999, MNRAS, 309, 253 

\bibitem[\protect\citeauthoryear{King, Taam, 
\& Begelman}{2000}]{Kingetal00} King A.~R., Taam R.~E., Begelman M.~C., 2000, ApJ, 530, L25 

\bibitem[\protect\citeauthoryear{Klu{\'z}niak 
\& Lasota}{2015}]{KL} Klu{\'z}niak W., Lasota J.-P., 2015, MNRAS, 448, L43 (KL)

\bibitem[Lasota et al.(2016)]{meOlek15} Lasota, J.-P., Vieira, 
R.~S.~S., Sadowski, A., Narayan, R., 
\& Abramowicz, M.~A.\ 2016, in press arXiv:1510.09152 

\bibitem[\protect\citeauthoryear{Mainieri et 
al.}{2010}]{Mainierietal10} Mainieri V., et al., 2010, A\&A, 514, A85 

\bibitem[Podsiadlowski 
\& Rappaport(2000)]{PR00} Podsiadlowski, P., \& Rappaport, S.\ 2000, ApJ, 529, 946 

\bibitem[Radhakrishnan 
\& Srinivasan(1982)]{RadSri82} Radhakrishnan, V., \& Srinivasan, G.\ 1982, Current Science, 51, 1096 

\bibitem[\protect\citeauthoryear{Ruderman 
\& Sutherland}{1975}]{RS75} Ruderman M.~A., Sutherland P.~G., 1975, ApJ, 196, 51 

\bibitem[Sadowski et al.(2016)]{Oleketal15} Sadowski, A., Lasota, 
J.-P., Abramowicz, M.~A., \& Narayan, R.\ 2016, MNRAS, in press; arXiv:1510.08845 

\bibitem[\protect\citeauthoryear{Shakura 
\& Sunyaev}{1973}]{SS73} Shakura N.~I., Sunyaev R.~A., 1973, A\&A, 24, 337 

\bibitem[Taam 
\& van den Heuvel(1986)]{Taamvdh86} Taam, R.~E., \& van den Heuvel, E.~P.~J.\ 1986, ApJ, 305, 235

\bibitem[Tong(2015)]{Tong15} Tong, H.\ 2015, Research in 
Astronomy and Astrophysics, 15, 517 

\bibitem[\protect\citeauthoryear{Webbink}{1977}]{W77}Webbink  R. F., 1977, ApJ, 211, 486

\bibitem[Wiktorowicz et al.(2015)]{Wiktor15} Wiktorowicz, G., 
Sobolewska, M., S{\c a}dowski, A., \& Belczy\'nski, K.\ 2015, ApJ, 810, 20 

\bibitem[\protect\citeauthoryear{Yong \& Li}{2015}]{YL15} Yong, S. \& Li, X.-D. \ 2015, ApJ, 802, 131


\end{thebibliography}
\end{document}